\documentclass{article}

\usepackage[nonatbib, preprint]{neurips_2025}

\usepackage[backend=bibtex, style=numeric]{biblatex}
\usepackage[utf8]{inputenc} 
\usepackage[T1]{fontenc}    
\usepackage{hyperref}       
\usepackage{url}            
\usepackage{booktabs}       
\usepackage{amsfonts}       
\usepackage{nicefrac}       
\usepackage{microtype}      
\usepackage{xcolor}         
\usepackage{graphicx}
\usepackage{multirow}
\usepackage{array}
\usepackage{longtable}
\usepackage{pdflscape}
\usepackage{float}
\usepackage{dblfloatfix}
\usepackage{caption}
\usepackage{enumitem}

\hypersetup{colorlinks=true, linkcolor=blue, urlcolor=blue}

\newcolumntype{L}[1]{>{\raggedright\let\newline\\\arraybackslash\hspace{0pt}}m{#1}}
\newcolumntype{C}[1]{>{\centering\let\newline\\\arraybackslash\hspace{0pt}}m{#1}}
\newcolumntype{R}[1]{>{\raggedleft\let\newline\\\arraybackslash\hspace{0pt}}m{#1}}

\title{TEDI: Trustworthy and Ethical Dataset Indicators to Analyze and Compare Dataset Documentation}

\author{Wiebke~Hutiri \quad Mircea~Cimpoi \quad Morgan~Scheuerman \quad Victoria~Matthews \quad Alice~Xiang\\ \\ Sony AI\\
}

\begin{document}

\maketitle

\begin{abstract}
Dataset transparency is a key enabler of responsible AI, but insights into multimodal dataset attributes that impact trustworthy and ethical aspects of AI applications remain scarce and are difficult to compare across datasets. To address this challenge, we introduce Trustworthy and Ethical Dataset Indicators (TEDI) that facilitate the systematic, empirical analysis of dataset documentation. TEDI encompasses 143 fine-grained indicators that characterize trustworthy and ethical attributes of multimodal datasets and their collection processes. The indicators are framed to extract verifiable information from dataset documentation. Using TEDI, we manually annotated and analyzed over 100 multimodal datasets that include human voices. We further annotated data sourcing, size, and modality details to gain insights into the factors that shape trustworthy and ethical dimensions across datasets. We find that only a select few datasets have documented attributes and practices pertaining to consent, privacy, and harmful content indicators. The extent to which these and other ethical indicators are addressed varies based on the data collection method, with documentation of datasets collected via crowdsourced and direct collection approaches being more likely to mention them. Scraping dominates scale at the cost of ethical indicators, but is not the only viable collection method. Our approach and empirical insights contribute to increasing dataset transparency along trustworthy and ethical dimensions and pave the way for automating the tedious task of extracting information from dataset documentation in future.

\end{abstract}

\section{Introduction}
\label{s:introduction}


Large-scale datasets are central to driving advances in AI~\cite{oala_dmlr_2023}, yet the collection of these datasets presents increasing ethical concerns~\cite{baack_training_2024, wyllie_fairness_2024}. For example, social biases~\cite{ross_measuring_2021}, gender biases~\cite{tang_mitigating_2021, booth_bias_2021}, sexual-objectification, and disparate emotion representation~\cite{wolfe_contrastive_2023} have been studied in vision-language models and found to be largely attributable to their training corpora~\cite{birhane_multimodal_2021, tang_mitigating_2021, wolfe_contrastive_2023}. Recent studies and dataset audits that have investigated specific harm areas like diversity and representation~\cite{zhao_position_2024, longpre_bridging_2025}, privacy~\cite{mittal_responsible_2024}, bias~\cite{shahbazi_representation_2023}, licensing~\cite{longpre_data_2023}, authenticity and consent~\cite{longpre_data_2024}, have found that concerns in these areas persist across datasets. While ethical concerns in text~\cite{elazar_whats_2024, longpre_data_2023, baack_training_2024, dodge_documenting_2021} and image~\cite{birhane_multimodal_2021, garcia_uncurated_2023, prabhu_large_2020, meister_gender_2023} datasets have been studied extensively, insights into datasets of other modalities like audio~\cite{longpre_bridging_2025}, music~\cite{morreale_data_2023} and speech~\cite{rusti_about_2023, leschanowsky_data_2025} remain scarce and oftentimes task specific. This leaves a gap in our understanding of how datasets amplify a broad spectrum of risks in multimodal models~\cite{liang_foundations_2024}. 


Dataset documentation, like datasheets~\cite{Gebru2021datasheets}, data cards~\cite{pushkarna_data_2022}, data statements~\cite{mcmillan-major_data_2024}, and crowdworksheets~\cite{diaz_crowdworksheets_2022} present an opportunity to gain insights into the ethical attributes of multimodal datasets. However, analyzing and comparing documentation across datasets is challenging. Different documentation approaches like datasheets and crowdworksheets focus on different ethical and trustworthiness aspects. This results in reported information that is overlapping between approaches, yet also has a high degree of variance~\cite{miceli_documenting_2021}. While more stringent legal requirements to publish training data summaries in templated formats suggests that regulators view dataset transparency as being in the public interest~\cite{noauthor_second_2025, noauthor_third_2025}, there is no standardized understanding of what makes a dataset ethical, and no global standard that mandates specific information that should be included in dataset documentation. Documentation is thus completed to varying degrees of consistency~\cite{yang_navigating_2024}. From a user perspective it can be time consuming and cumbersome to extract information from documentation, as it is typically created in natural language text. Even though metadata standards have recently been proposed for machine learning datasets~\cite{akhtar_croissant_2024} and extended to responsible AI~\cite{jain_standardized_2024, roman_open_2024}, they are not yet widely adopted and will require extensive effort to apply to datasets released in the past~\cite{bandy_addressing_2021}. 

Beyond studying trustworthy and ethical dataset attributes, there is a growing need to understand how these attributes arise from design choices in sociotechnical AI systems, so that leverage points for intervening mitigations can be identified. For example, recent research has drawn attention to the irreconcilable tension between scraping and privacy laws~\cite{solove_great_2024}, pointing to the influence that this specific data collection method has on the ethical attributes of datasets. Similarly, several dataset studies have identified collection methods and sources for datasets~\cite{andrews_ethical_2023, morreale_data_2023, longpre_bridging_2025}, implying that ethical attributes are not coincidental, but rather arise from sociotechnical factors and developer choices around how data is collected. However, empirical studies that systematically interrogate the impact of functional aspects of datasets, like size, modality, and collection methods, on a broad set of ethical indicators are currently lacking.

This paper sets out to develop a systematic approach to extract reliable, verifiable information about trustworthy and ethical attributes of datasets from their documentation. In doing this, we aim to leverage documentation to identify attributes of individual datasets, to compare datasets, and to study sociotechnical factors and design choices that impact trustworthy and ethical indicators. We contribute to ongoing research efforts that aim to improve dataset transparency by: 

\begin{enumerate}[noitemsep]
    \item Presenting Trustworthy and Ethical Dataset Indicators (TEDI) as a framework for the verifiable assessment of dataset documentation.
    \item Auditing documentation of multimodal datasets of underexplored modalities along a broad spectrum of ethical indicators using TEDI.
    \item Taxonomizing data collection methods, and studying how they impact trustworthy and ethical dimensions of datasets.
\end{enumerate}

\section{Trustworthy and Ethical Dataset Indicators (TEDI)}
\label{s:tedi}



TEDI\footnote{TEDI can be accessed \href{https://docs.google.com/spreadsheets/d/e/2PACX-1vSGzWJzfnZ0jUeyByGWGmZimpwdqk-89YPAq_oKfyjkO2w4Gai-eDuSAqvMF3CHvI7AE7nd7IHZkA-R/pubhtml?gid=0&single=true}{online}.} is a systematic framework to assess aspects of AI data collections that affect the ethics and trustworthiness of AI systems built on them. Particularly, TEDI leverages dataset documentation, including research papers, datasheets, crowdworksheets, and blog posts or websites created by the dataset authors or their organisation, as evidence to capture attributes of datasets and their collection processes. By capturing the rich information contained in dataset documentation in a standardised format, TEDI converts dataset documentation from a reporting tool into an information source suitable for analyzing and comparing datasets across trustworthy and ethical dimensions. TEDI categorizes 143 fine-grained trustworthiness and ethical concerns in a 3-level hierarchy. 
The hierarchy was developed iteratively, using a design science research approach~\cite{hevner_design_2010}. We started with categories derived from literature and theory, and added additional categories if new indicators could not be labelled readily as belonging to an existing category. 

At the top level TEDI indicators are divided into \textit{ethical} and \textit{trustworthiness} categories. This differentiation reflects the distinction drawn in the EU Trustworthy AI Guidelines~\cite{noauthor_ethics_2019} between ethical principles that form the foundations of Trustworthy AI and the requirements that are necessary to realise Trustworthy AI. Similarly, we position that ethical dataset indicators need trustworthiness indicators to support their realization. At the second level, the ethical categories are grounded in the Belmont Report~\cite{noauthor_belmont_1979} and its expanded four principles of research ethics~\cite{beauchamp_principles_2013}, namely \textit{autonomy}, \textit{beneficence}, \textit{justice}, and \textit{no-harm} (i.e. non-maleficience). The supporting trustworthiness categories draw on the EU Trustworthy AI Guidelines, and are \textit{reliability}, \textit{transparency} and \textit{accountability}. At the third level, categories are broken down into further subcategories that extend the concepts present at the second level. For example, indicators that are classified as impacting \textit{autonomy} can be further divided into \textit{consent} and \textit{control}. Finally, each indicator is captured as a short phrase or statement that represents a single idea, such as \textit{data subjects provided consent} or \textit{voluntary consent}. The majority of indicators are drawn from existing documentation approaches~\cite{Gebru2021datasheets, chmielinski_dataset_2022, bommasani_foundation_2023, diaz_crowdworksheets_2022}, dataset measurement~\cite{mitchell_measuring_2023, zhao_position_2024}, and best practices for data curation from literature and industry guidelines~\cite{andrews_ethical_2023, noauthor_improving_nodate, noauthor_data_2023}. They were validated using informal interviews with researchers and practitioners who have first-hand experience with data collection. 

To turn TEDI into a verifiable assessment rubric, each indicator is rephrased as a question that targets the information contained in dataset documentation, for example the \textit{data subjects provided consent} indicator is phrased as \textit{"Does the document mention that data subjects provided consent?"}. The categorised questions are listed in Table~\ref{tab:tedi_detail} in the Appendix. TEDI's questions proposition that transparency is a necessary precondition for judging the trustworthy and ethical attributes of a dataset. However, statements are more credible if they are backed up with evidence or justified. To account for this, questions can be answered with a limited set of four responses: \textit{not applicable}, \textit{no}, \textit{yes}, or \textit{yes with evidence or justification}. Questions are formulated to enforce a consistent response interpretation, so that `yes' presents a positive signal for the indicator, while `no' indicates the absence of information and presents no signal. It is important to note that a positive signal for an indicator relates to the availability of information in the dataset documentation and is not a confirmation that the dataset itself meets a particular ethical standard implied by the indicator. To move from the evidence provided in the documentation to a judgement on the dataset itself, a subjective assessment by an auditor will typically be necessary. 

To illustrate the difference between response options, consider the following example. Documentation may mention that the provision of consent was voluntary, leading to a `yes' response to the \textit{voluntary consent} indicator. If the documentation then continued to provide a sample of the messaging that data subjects received, the indicator response would be adjusted to `yes with evidence or justification'. In an ideal case, all applicable questions for a dataset should be answered with `yes with evidence or justification'. As a next step, an auditor may analyze the message to data subjects, passing the judgement that data subjects had the freedom not to participate and that the \textit{voluntary consent} indicator has indeed been met for the dataset. This subjective assessments about the dataset, as opposed to its documentation, is desirable but falls outside the scope of TEDI. 



\section{Analysis of Multimodal Datasets}
\label{s:method}

Datasets that include people heighten certain risks related to privacy, personal information, consent and bias~\cite{andrews_ethical_2023, xiang_being_2022}. As far as ethical considerations are concerned, human-centric datasets thus tend to be viewed with greater scrutiny than those that do not include people. Despite the human voice being a biometric marker that can reveal not only your identity, but also your health status and emotional states~\cite{hansen_speaker_2015}, only few studies have investigated ethical aspects of datasets that include the human voice~\cite{longpre_bridging_2025, rusti_about_2023}.
With the increasing popularity of multimodal, voice-based interfaces that rely on speech data, filling this gap is important. To demonstrate the utility of TEDI, we thus audit human-centric multimodal datasets that include human voices either directly, e.g. speech datasets that explicitly collected spoken words, or indirectly, e.g. video datasets that include singing, talking or other human-made sounds.

\subsection{Dataset Selection and Annotation}

An overview of our dataset selection approach is shown in Figure~\ref{fig:method} in the Appendix. We selected all datasets on PapersWithCode that matched on keywords and task-specific search terms. Datasets that did not include the human voice were removed. Additionally, we only included datasets with more than 10 Google Scholar citations on 30 August 2024. This selection process resulted in 77 datasets. Next we adopted a snowball sampling approach, where we included datasets mentioned in the related works section or the utility evaluations of the datasets selected in the previous step. Finally we conducted purposive sampling of speech datasets that were introduced in recent publications that did not yet meet the citation requirements or that were not on PapersWithCode (e.g. Libriheavy~\cite{kang_libriheavy_2024} and Yodas~\cite{li_yodas_2024}). Many text-to-speech datasets are poorly or not at all documented and not citable, so we relied on the authors' expertise to select frequently used datasets in this task domain. We ultimately ended up with 114 datasets for analysis\footnote{The annotated datasets are accessible \href{https://docs.google.com/spreadsheets/d/e/2PACX-1vRJm-0VFJxCpcGwLpreN1wbqcjrElX4DFFnEnXMgDEeGSFmqPhkDE9J7tfkFz-u_tYvaHjQXAidPYou/pubhtml}{online}.}. For each dataset, we annotated functional attributes related to the type of instances (e.g. video clip, speech recording) and dataset size (i.e. number and duration of instances). Furthermore, we annotated datasets along ethical and trustworthiness dimensions using TEDI and labelled data sourcing information (to be discussed in detail in Section~\ref{s:results}). 


\subsection{The Anatomy of a Multimodal Dataset}

By definition, multimodal datasets combine multiple modalities, which can be collected in different ways and have different trustworthy and ethical attributes. To enable insights into multimodality and data sourcing, we introduce a convention for disambiguating primary, secondary, and annotation modalities of datasets. We define the primary modality as the modality of the data source, or as the modality driving the task for which the dataset was created. The secondary modality is then an additional modality collected alongside, extracted from, or accompanying the primary modality. Annotations are additional modalities that are provided after the collection of the primary or secondary modality. With this convention, multimodal datasets can be categorized by their pairings across modalities. Table~\ref{tab:dataset_modalities} lists the most common modality pairs for Primary-Secondary and Primary-Annotation modalities, an example that details the typical instances in these pairs, and the frequency of occurrence of the pairs in the corpus of datasets we analyzed. Video data is the dominant primary modality -- perhaps unsurprisingly, given the abundance of video data on the web, and the frequency of data collection via scraping, as we will show in Section~\ref{s:results}. The most commonly occurring Primary-Secondary modality pairs are video-speech and speech-text data. As expected, text data is the dominant annotation modality. Non-text annotations are mostly pixel-level annotations, such as bounding boxes and segmentation masks, or datasets without annotations.

Figure \ref{fig:hist_data_detail} in the Appendix further disaggregates speech, video, and text modalities by detailed modality descriptions. We use the figure to analyze the coverage of data modalities in primary and secondary data sources. For datasets where speech is the primary modality, read speech dominates the type of speech that has been collected. When speech is the secondary modality, emotive and spontaneous speech have been prioritized in datasets. A reason for this coverage of speech data may be that read speech can be easily derived from audio recordings, particularly from audiobooks. However, more natural forms of speech are difficult to collect and are typically extracted from video data. Popular types of video data are talking heads, face clips, and actions, but many datasets also collect videos indiscriminately, without specifying the type of content that is sought out. Scripts, labels, transcripts, and metadata are popular text annotations. Further text annotations that are frequently collected include normalized text, which has been processed to meet the requirements of automatic speech recognition (ASR), and alt-text extracted from web sources. Scripted text is also a frequent secondary modality, as it forms the underlying content of read-speech corpora that are critical for ASR.

\begin{table}[bth]
  \centering
  \footnotesize
    \caption{Common modality pairs for Primary-Secondary and Primary-Annotation modalities in the datasets that we analyzed.}
    \label{tab:dataset_modalities}
    \begin{tabular}{lllc}
    \toprule
    & \textbf{Modality Pair} & \textbf{Example} & \textbf{Occurrence} \\ \midrule
    \multirow{4}{*}{Primary-Secondary} & video-speech & YouTube video of a lecture & 37\%\\
       & speech-text & audio book recording with the text of the book & 17\%\\
       & video-audio & 360 degree audio with environment sounds & 10\%\\ 
       & video-text & TV show with subtitles & 6\%\\
       & other & - & 30\% \\ \midrule
    \multirow{4}{*}{Primary-Annotation} & video-text & video with description & 52\% \\
       & speech-text & audio book recording with transcriptions & 19\%\\
       & image-text & photo with caption & 8\%\\
       & audio-text & audio recording with caption & 3\%\\
       & other & - & 17\%
    \end{tabular}
\end{table}

\subsection{Dataset Usage}

\begin{figure}[hbt]
    \centering
    \includegraphics[width=\linewidth]{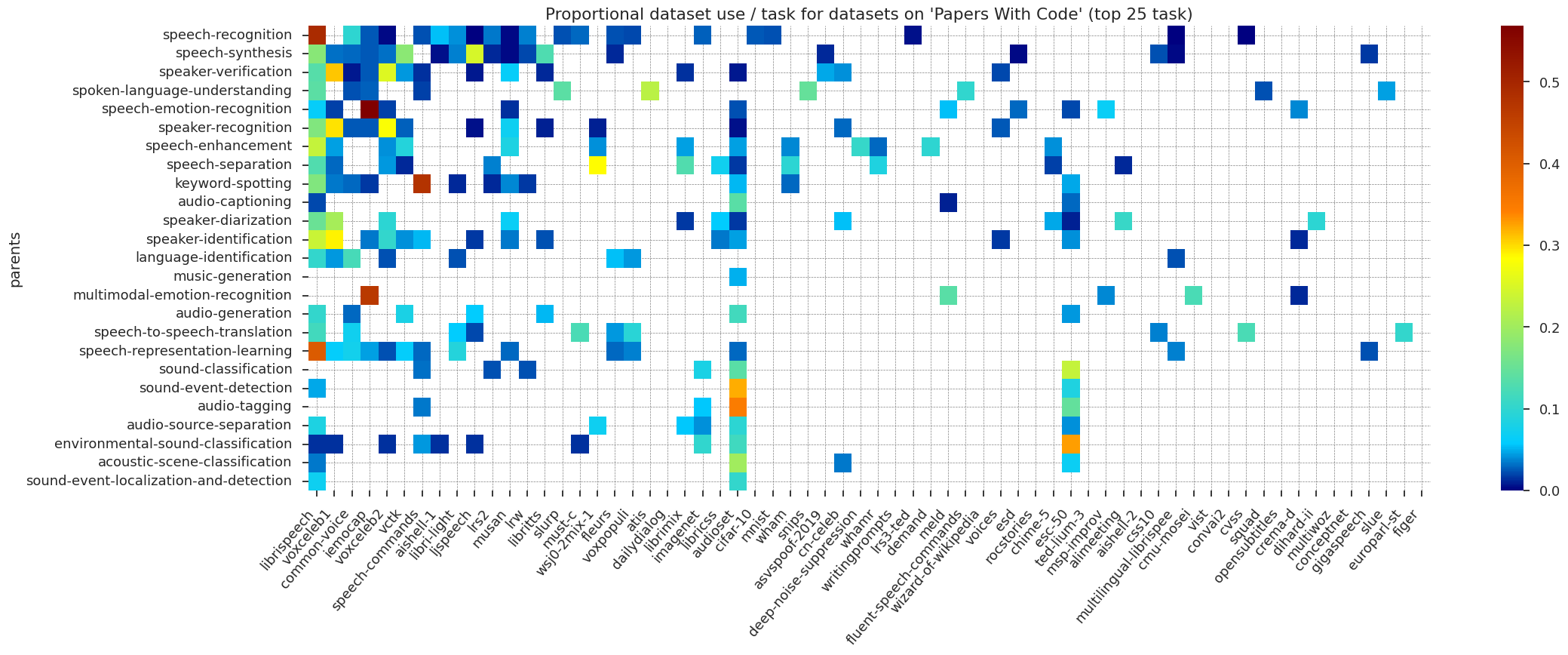}
    \caption{Proportional dataset use per task for datasets from PapersWithCode}
    \label{fig:pwc_speech_dataset_count}
\end{figure}

Many multimodal datasets that include speech tend to be task-specific and a select few datasets dominate individual tasks. For datasets that are on PapersWithCode, the heatmap in Figure~\ref{fig:pwc_speech_dataset_count} visualizes how frequently datasets are used for specific tasks, based on their use in papers on the platform (see Figure~\ref{fig:pwc_paper_count} in the Appendix for the distribution of papers across tasks). The most notable dataset is Librispeech~\cite{Panayotov2015}, which is heavily used across a wide variety of tasks. It is also used in half the papers that focus on ASR, the most dominant speech processing task. Given its popularity and wide-scale use, Librispeech can be considered the ImageNet~\cite{deng_imagenet_2009} of speech processing. Similarly, Audioset~\cite{gemmeke_audio_2017} dominates audio processing tasks, in particular audio tagging and sound event detection. VoxCeleb1~\cite{nagrani_voxceleb_2017} and 2~\cite{chung_voxceleb2_2018} are the most prominent datasets in speaker recognition research, including speaker identification and speaker verification. IEMOCAP~\cite{busso_scripted_2008}, a video-speech dataset with dense text annotations, remains a key data resource for speech and multimodal emotion recognition, with the more recent MELD~\cite{poria_meld_2019} dataset also showing high usage patterns.

Mirroring the trends that neural networks have spurred across the board, research progress in speech and multimodal data processing over the past decade has been driven by the collection of ever larger datasets. Figure~\ref{fig:release_count_hours} in the Appendix shows a histogram of the release years of datasets in this study, overlaid with the mean dataset size in hours for each year. Prior to 2014, datasets were small, containing up to 100 hours of speech data. 2014 saw the release of ActivityNet~\cite{heilbron_activitynet_2015} and YFCC100M~\cite{thomee_yfcc100m_2016} as large-scale video-text and image-text datasets. Following this, in 2015, Librispeech scaled the speech-text modality to 982~hrs. Since then new datasets have been released continuously. ASR in particular has pushed the frontiers of large scale speech-text data collection, with Libri-Light~\cite{kahn_libri-light_2020} (62.2K hrs) scaling speech datasets from thousands to ten-thousands of hours in 2019, and Multilingual Librispeech~\cite{pratap_mls_2020} (50,834~hrs) and Common Voice~\cite{ardila_common_2020} (32,121~hrs) scaling from English to multiple languages in 2020. Following this trend, Gigaspeech~\cite{chen_gigaspeech_2021} was released in 2021 with 10K~hrs, and VoxPopuli~\cite{wang_voxpopuli_2021} with 400K~hrs of speech data. Despite the trend for scale driven by ASR, small and bespoke datasets, such as IEMOCAP (12hrs, released in 2008), VCTK~\cite{yamagishi_cstr_2019} (first released in 2012) and LJSpeech~\cite{noauthor_lj_nodate}(23.9hrs, released in 2017) continue to have a large influence on important tasks such as emotion recognition and speech synthesis. 



\section{Impact of Data Sourcing on Trustworthy and Ethical Indicators}
\label{s:results}

Data collection methods constrain how data is collected, determine the data source, and influence the quantity of data that can be collected. Data sourcing -- which includes methods used to collect data and the sources that data is collected from -- is thus poised to wield influence over trustworthy and ethical dataset indicators by shaping processes and dataset attributes. In this section, we present a novel, granular categorization scheme of data collection methods, which we used to annotate our corpus of datasets. We then show how requirements for dataset scaling influence possible data collection methods and how this ultimately impacts the extent to which the indicators are documented.  

\subsection{Data Sourcing Taxonomy}

To ground our categorization of data collection methods, we created a two-tier data sourcing taxonomy (see Table~\ref{tab:data-collection-taxonomy}). We developed the taxonomy through iteration, starting with literature and then augmenting the taxonomy with collection methods we identified in our dataset corpus. The high level \textit{Collection Method} in our taxonomy is based on the categorisation scheme proposed in \cite{zhao_position_2024}. We expanded their five categories to seven categories, adding \textit{Sampling} and \textit{Proprietary} collection methods when we observed datasets that used these methods. In addition to the seven top-tier categories, we include 25 sub-categories. Table~\ref{tab:data_collection_tax_long} in the Appendix lists all sub-categories, their definitions, and an example dataset for each. The sub-categories for the \textit{Direct Collection} method apply either to annotations (self-reported, expert, amateur) or data samples (lab, wild, constrained). Modalities can be sourced with multiple collection methods. For example, the primary and secondary modalities of Ego4D~\cite{grauman_ego4d_2022} were sourced with in-the-wild direct collection and paid crowdsourcing. We used the taxonomy to annotate the data collection methods for the primary, secondary, and annotation modalities of all datasets. In addition to the data collection methods, we also identified from where data was sourced. Where possible, we further identified the origin of the data source -- that is to say, the types of entities or individuals contained in the source.

\begin{table}[hbt]
  \caption{Data Sourcing Taxonomy with high level collection methods, definitions and sub-categories. Asterisks indicate definitions *similar to or **adopted from~\cite{zhao_position_2024}}
  \label{tab:data-collection-taxonomy}
  \centering
  \scriptsize
  \renewcommand{\arraystretch}{1.35}
  \begin{tabular}{L{0.14\textwidth}L{0.51\textwidth}L{0.24\textwidth}}
    \toprule
    \textbf{Collection Method} & \textbf{Definition} & \textbf{Sub-categories} \\ \midrule 
    \textbf{Sampling} & Instances are selected using an implicit or explicit sampling strategy. & convenience, purposive \\
    \textbf{Direct Collection} & Instances are collected directly from, or annotated by, a human whose identity is known to the research team or data collectors. & a: self-reported, expert, amateur, d: lab, wild, constrained \\
    \textbf{Proprietary} & Instances are owned by the data collectors, or data collectors believe they are their lawful custodian. & internal, agreement, purchased \\
    \textbf{Crowdsourced} & Instances are collected from or with the help of a group of typically anonymous data workers or people making data donations who are not directly known to the research team or data collectors.* & volunteer, compensated, paid \\
    \textbf{Scraped} & Instances are retrieved in a large volume from the Internet using routine or automatic means.* & crawled, sources with curation, constrained \\
    \textbf{Derived} & Instances are sourced from a previously existing data collection.** & extracted, copy, subset, extended, expanded, composition \\
    \textbf{Synthetic} & Instances are artificially manufactured or procedurally generated rather than capturing real-world events.** & extended, expanded \\
    \bottomrule
  \end{tabular}
\end{table}

\subsection{Disentangling the Data Collection Web}

The dominant collection methods for primary modalities are scraped, derived, and direct collection (see Table~\ref{tab:data-collection-methods}). Over half the secondary modalities were derived, while most other data were obtained via direct collection or sampling. Meanwhile, a third of the annotations in our corpus were derived. Annotations sourced via synthetic means, crowdsourcing, and direct collection were also frequent. To drill deeper, we use Sankey diagrams (included in the Appendix) to explore trends in data collection methods and sources between primary, secondary, and annotation modalities. Primary modalities (Figure~\ref{fig:sankey_primary}) that have been scraped are dominated by video data from YouTube. Derived primary modalities are mostly subsets of existing data collections, with the LibriVox project being the most favoured parent data source. Direct collection is mostly done in the lab. For several datasets, actors were recruited to participate in the data collection -- a practice that is most common for the collection of emotive data. 

Derived secondary modalities (Figure~\ref{fig:sankey_secondary}) are typically speech data that have been extracted from a  video data source (the primary modality). Direct collection of secondary modalities almost always co-occurs with direct collection of primary modalities, and is thus is done in lab settings. Sampling mostly occurs when the secondary modality is text and the primary modality is speech. Purposive sampling approaches are oftentimes used here, for example when specific books or audiobook projects (Project Gutenberg and Aozora Bunko) are selected. As with secondary modality data, derived annotations are typically extracted from the primary data source. Synthetic extensions leverage a wide variety of automatic pipelines and AI systems, primarily to extract text from speech with automatic speech recognition. The majority of crowdsourced annotations have been provided by paid gig workers on Amazon Mechanical Turk.

\begin{table}[hbt]
  \caption{Proportional data collection methods for primary, secondary and annotation modalities}
  \label{tab:data-collection-methods}
  \centering
  \small
  \renewcommand{\arraystretch}{1.05}
    \begin{tabular}{lccc}
    \toprule
         & \textbf{Primary} & \textbf{Secondary} & \textbf{Annotation}\\\midrule 
        \textbf{Sampling} & 8\% & 17\% & 13\%\\
        \textbf{Direct collection} & 23\% & 20\% & 18\%\\
        \textbf{Proprietary} & 7\% & 2\% & -\\
        \textbf{Crowdsourced} & 11\% & 5\% & 23\%\\
        \textbf{Scraped} & 30\% & 5\% & -\\
        \textbf{Derived} & 25\% & 52\% & 36\%\\
        \textbf{Synthetic} & 2\% & 3\% & 28\%\\
        \textbf{Unspecified} & - & - & 12\% \\
    \bottomrule
    \end{tabular}
\end{table}

\begin{figure}[hbt]
    \centering
    \includegraphics[width=0.9\linewidth]{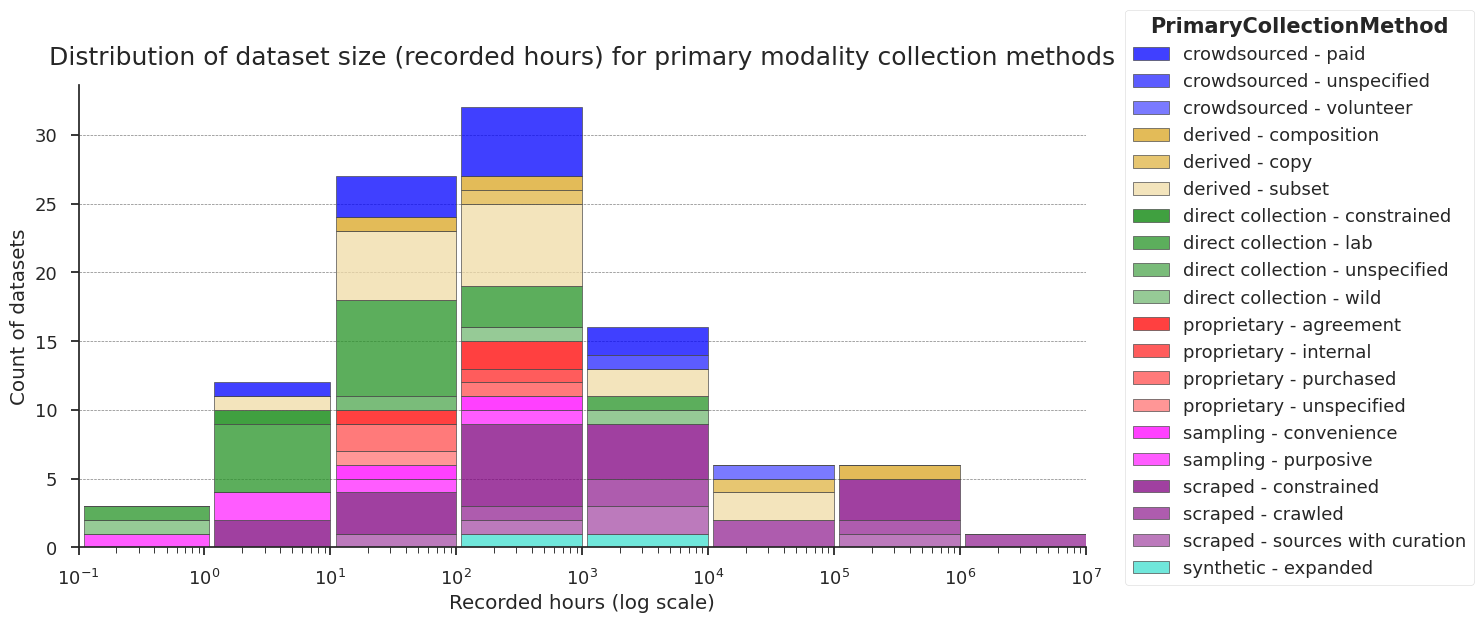}
    \caption{Data collection methods influence the size (i.e. recorded hours) of datasets.}
    \label{fig:histogram_primary_cmsize}
\end{figure}

Data collection methods are closely related to the quantity of data that has been collected. As shown in Figure~\ref{fig:histogram_primary_cmsize}, small datasets up to 10~hrs are mostly collected via direct collection. Between 10 and 1K~hrs, a wide variety of collection methods are at play. Nonetheless, direct collection remains important for smaller datasets. As dataset size increases from 100 to 1K~hrs, scraped, derived, and crowdsourced methods become more prominent. Beyond 1K~hrs, scraped datasets dominate. However, there are some datasets that do not fit these trends. Golos~\cite{karpov_golos_2021} (1.24K~hrs) and Ego4D~\cite{grauman_ego4d_2022} (3.67K hrs) have been collected with mixed methods, using both direct collection and paid crowdsourcing. For Golos, only 10\% of data was collected via direct collection, with the remainder collected via a crowdsourcing platform. For Ego4D, the exact split between the collection methods is difficult to determine. FLEURS~\cite{conneau_fleurs_2022} (1.4K~hrs), a multilingual read speech dataset, is the only other dataset in the 1K - 10K range in our corpus where the primary modality has been collected via crowdsourcing. The three derived datasets above 10K~hrs, Libri-light~\cite{kahn_libri-light_2020}, Libriheavy~\cite{kang_libriheavy_2024} and Multilingual Librispeech~\cite{pratap_mls_2020}, are all subsets of the LibriVox project with over 50K~hrs in duration. As far as collection methods go, Common Voice~\cite{ardila_common_2020} stands apart from the other datasets in our corpus. It is an ongoing data collection that contains over 30K~hrs of crowdsourced data from volunteers. 

\subsection{How Data Collection Methods Impact Trustworthy and Ethical Dataset Indicators}
\label{ss:methods_impact_indicators}

We now use TEDI to analyze how trustworthy and ethical dataset indicators are accounted for in our dataset corpus and different data collection methods. Given findings from prior literature, our base assumption was that many datasets were unlikely to have considered or documented trustworthy and ethical indicators in detail. We thus decided to use TEDI's third-level categories for a high-level assessment of the datasets and applied the following rules during annotation. For each third level category, if the documentation allowed us to respond `yes' to any of the indicators associated with that category, we marked that category as `yes'. For example, the \textit{consent} category includes nine indicators. If documentation satisfied any one of those indicators, we annotated the dataset with `yes' for \textit{consent}. Our analysis thus presents the most optimistic perspective on the extent to which TEDI are covered in dataset documentation.

\begin{figure}[bth]
    \centering
    \includegraphics[width=0.85\linewidth]{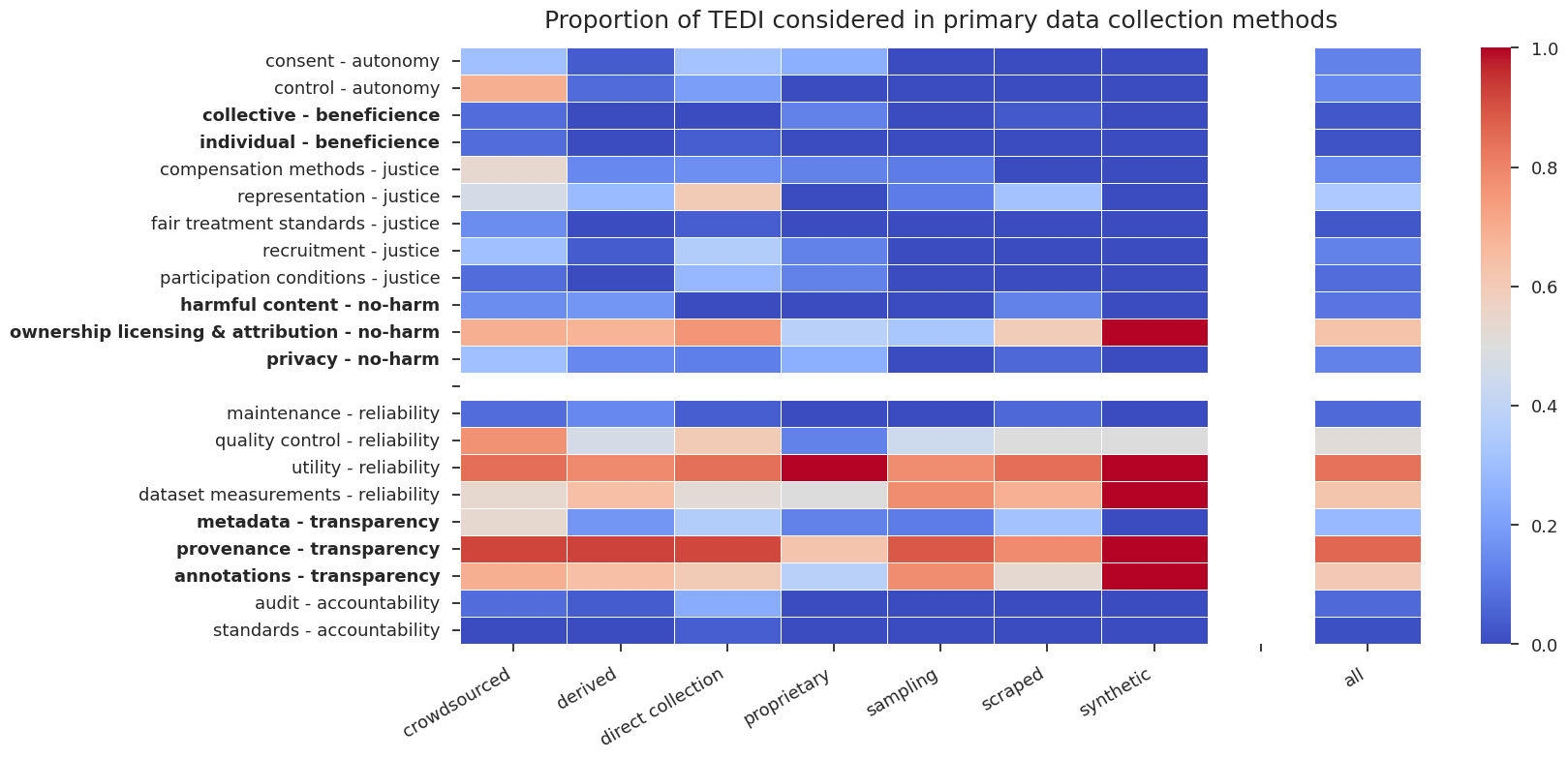}
    \caption{Proportion of datasets that have considered trustworthy (bottom) and ethical (top) dataset indicators for primary data collection methods.}
    \label{fig:eti_collection_methods}
\end{figure}

Figure~\ref{fig:eti_collection_methods} shows the proportion of datasets that have considered trustworthy and ethical indicators for particular collection methods and for all datasets combined (right). It is clear that trustworthiness indicators are considered more frequently than ethical indicators. In particular, \textit{utility}, which includes indicators for task evaluation, cross-dataset generalization, variance, and prediction confidence, was documented in the vast majority of datasets in our corpus. This is unsurprising, as datasets are often released to introduce a new algorithmic technique or architecture, which presupposes a utility evaluation. With the exception of proprietary datasets, the majority of datasets also considered \textit{provenance} related indicators to some extent. Amongst transparency indicators, the lack of \textit{metadata}, which includes basic considerations such as defining variables and variable data types, is concerning. In addition, very few datasets considered basic \textit{maintenance} related aspects, namely establishing a process for data record removal, versioning and dataset deprecation. Accountability categories were rarely mentioned in dataset documentation, with only a select few direct collection datasets having conducted an independent ethics review or implemented external standards.

Amongst ethical indicators, \textit{ownership, licensing, and attribution}, is the only category that was considered by more than half the datasets in our corpus. This category had a low barrier to entry in our study, as merely mentioning a dataset license would result in a positive assessment. For the remaining ethical indicators, considerations were low and varied based on collection methods. Datasets that have been sampled, scraped, or created via synthetic means had almost no further mention of information related to any ethical indicators in their documentation. Crowdsourced and direct data collections were more likely to consider ethical indicators than other collection methods. Over half the crowdsourced collections considered indicators that relate to participant or data subject \textit{control}, such as pilot data collections, self-reported variables, and the incorporation of data subject feedback in the collection task design. \textit{Compensation methods} were mostly limited to a mention of whether participants were compensated for data work, and were reported by just over half the crowdsourced datasets. \textit{Representation} related indicators were mentioned in more than half the direct collections and in just under half the crowdsourced collections. Despite increasing attention to the importance of consent in global data protection regulations, \textit{privacy} and \textit{consent} were largely ignored and remained conspicuously absent from scraped and derived datasets.  
\section{Discussion and Limitations}
\label{s:discussion}

This paper proposes Trustworthy and Ethical Dataset Indicators (TEDI) as a verifiable rubric to extract information from dataset documentation. Prior research has suggested approaches to improve dataset documentation with machine-readable metadata~\cite{akhtar_croissant_2024}, which have been extended to cover responsible AI. The Croissant-RAI schema~\cite{jain_standardized_2024} includes 20 responsible AI properties for 5 use cases: data life cycle, regulatory compliance, data labelling, AI safety and fairness, and participatory data. The properties of Croissant-RAI primarily cover indicators in our trustworthiness category, and rely on text descriptions. The Open Datasheets~\cite{roman_open_2024} metadata framework comprises general metadata properties and responsible AI metadata. It provides a convenient way of responding to questions outlined in Datasheets for Datasets~\cite{Gebru2021datasheets}, but also relies on text descriptions. These schemas are limited in scope, and it remains cumbersome to extract specific details about ethical categories like consent, control, compensation and harmful content from them. TEDI goes beyond the metadata properties established in these prior works to present 143 indicators, categorized in a 3-level hierarchy that is grounded in established frameworks for research ethics and trustworthy AI. Collectively the indicators present a holistic and systematic approach to capture trustworthy and ethical dimensions of datasets reported in documentation. The indicators provide evidence to answer the question \textit{"what makes one dataset more ethical than another dataset?"}. Importantly, they are structured to provide verifiable responses with consistent interpretation that can be extracted from dataset documentation. Through its systematic approach, TEDI bridges the gap between documentation that is difficult to analyze and compare, and machine readable metadata for responsible AI.

Despite regulatory calls for transparency in AI pipelines~\cite{noauthor_regulation_2024, noauthor_ab-2013_2024, translate_interim_2023}, ongoing dataset transparency concerns continue to be highlighted in the literature~\cite{longpre_data_2023, longpre_data_2024, longpre_bridging_2025, yang_navigating_2024, bommasani_foundation_2023}. While prior large scale audits have provided insights into specific attributes of datasets, such as licensing~\cite{longpre_data_2023}, harmful content~\cite{birhane_multimodal_2021} and bias~\cite{garcia_uncurated_2023}, some attributes, like consent and control, compensation, and privacy are difficult to audit at scale. 
Using TEDI, we contribute to ongoing research on dataset~\cite{prabhu_large_2020, zhao_position_2024, meister_gender_2023, elazar_whats_2024} and documentation~\cite{yang_navigating_2024, miceli_documenting_2021} audits, with a focus on the underexplored space of multimodal datasets that include human voices. Prior research~\cite{longpre_bridging_2025} has studied multimodal datasets, including 95 speech datasets that focus on automatic speech recognition (ASR). The study investigated licensing, geographic and linguistic diversity, data sources and collection methods. Our analysis presents a complementary perspective to that work, auditing datasets for tasks beyond ASR and considering the broad set of trustworthy and ethical dimensions captured by TEDI. We found that while some indicators are frequently reported on -- most notably indicators relating to annotations, provenance, utility, dataset measurement and licensing -- most are absent from dataset documentation. However, even categories with greater reporting progress have a long way to go to improve the state of trustworthy and ethical considerations in data collections. Our category-level dataset coding (described in Section~\ref{ss:methods_impact_indicators}) leads to an analysis that paints the most optimistic view of a poor state of affairs. For example, the mere existence of a dataset license does not preclude ownership issues with the data. As shown in previous work, dataset licenses, even if provided, are often inconsistent and do not comply with upstream licenses, or licenses of individual data records~\cite{longpre_data_2023, longpre_bridging_2025}. 



TEDI should not be considered complete. There remains scope for future work to analyze further datasets, and to expand on TEDI's indicators as required. Follow-on work should use TEDI to conduct a fine-grained analysis of datasets using the 143 indicators. However, annotating dataset documentation retrospectively is a time-consuming undertaking. Our approach of framing indicators as verifiable questions lends itself to automation and information retrieval at scale, alleviating the effort required to do this. It remains an area of future work to investigate how best to do this, and how best to integrate TEDI with existing responsible metadata schemas such as Croissant-RAI. We have demonstrated the use of TEDI for analyzing dataset documentation, but believe that TEDI can also support auditing processes and data collection design. Future work should study how auditors can use TEDI to deliver consistent insights into datasets, and how dataset creators can use TEDI to pre-emptively consider trustworthy and ethical dataset indicators in the design of new data collections. 

Our study only presents a partial view of the state of ethical considerations in multimodal data collections. While large, our corpus is limited by our selection and filtering criteria, e.g. we may have missed very recent datasets due to using citations as selection criteria. Our analysis also skews towards multimodal datasets that are used for speech processing, or that explicitly include speech. Many video data collections may include speech, but have historically not been used for speech processing if they had no ground truth text annotations. However, we expect that the majority of these data will be scraped from YouTube, and have similar attributes to the datasets we analyzed.




\section{Conclusion}
\label{s:conclusion}

Current dataset documentation approaches do not lend themselves readily to empirical analysis, making it difficult to compare datasets along ethical dimensions. We introduce Trustworthy and Ethical Dataset Indicators (TEDI) to address this challenge. TEDI encompasses 143 fine-grained indicators that can be extracted from dataset documentation. Each indicator is formulated as a question that yields a consistent and verifiable response, so that responses can be aggregated and analyzed. We also introduce a taxonomy of data collection methods to consider the impact of data sourcing practices on trustworthy and ethical dataset indicators. Using TEDI and the data sourcing taxonomy, we annotated and analyzed over 100 multimodal datasets. Our findings show how data collection methods impact the ethical attributes of datasets, and highlight that many multimodal datasets prevalent in research do not meet minimum ethical standards. There is significant opportunity to improve on all ethical categories in multimodal data collections, yet our analysis also highlights the tensions between dataset scale and ethical demands. Our verifiable assessment framework and findings pave the way for automating the tedious task of extracting information from dataset documentation in future.





{
\tiny
\printbibliography
}


\appendix
\newpage
\section{Appendix}

\subsection{Supporting Material: Trustworthy and Ethical Dataset Indicators}
\label{a:tedi}

\begin{scriptsize}
\setlength{\tabcolsep}{4pt}
\begin{longtable}{cL{0.25\textwidth}@{\hspace{2pt}}L{0.5\textwidth}} 
    \caption{Detailed Questions for the Trustworthy and Ethical Dataset Indicators}
    \label{tab:tedi_detail} \\
    \multirow{9}{*}{Consent} & Does the document mention that & data subjects provided consent?\\
    & Does the document mention that & consent is voluntary?\\
    & Does the document mention that & consent is informed?\\
    \textbf{Autonomy} & Does the document mention that & consent is meaningful?\\
    & Does the document mention if & consent revocation is possible (during data collection)?\\
    & Does the document mention if & consent revocation is possible after data collection?\\
    & Does the document mention if & non-consensual data subjects are removed from the dataset?\\
    & Does the document mention if & consent revocation requests have a SLA?\\
    & Does the document mention if & consent mechanisms align with context / cultural norms?\\ \midrule
    \multirow{9}{*}{Control} & Does the document explain why & the level of granularity of collected variables is justified? \\
    & Does the document mention if & the selection of multiple variable values was piloted?\\
    & Does the document mention if & open-ended response options were piloted?\\
    \textbf{Autonomy} & Does the document mention if & data subjects can influence what data samples are collected?\\
    & Does the document mention if & data subjects can influence metadata variables?\\
    & Does the document mention if & data subjects can influence metadata variable values?\\
    & Does the document mention if & data subjects can influence target labels?\\
    & Does the document mention & self-reported variables?\\
    & Does the document mention if & data subject feedback has been used for task improvement? \\ \midrule
    \multirow{2}{*}{\textbf{Beneficence}} & Does the document mention & specific benefit for data subjects?\\
     & Does the document mention & benefit that is grounded in reality?\\
    Individual / & Does the document mention & compelling justification for the benefit?\\
    Collective & Does the document mention & why the benefit is necessary?\\
    & Does the document mention & how the benefit is proportional to the data collection?\\ \midrule
    \multirow{9}{*}{Representation} & Does the document mention & data population statistics?\\
    & Does the document mention & data coverage across groups?\\
    & Does the document mention & the relationship between annotated variables and target variables?\\
    \textbf{Justice} & Does the document mention & the basis for selecting metadata variables?\\
    & Does the document mention & the basis for selecting metadata variable values?\\
    & Does the document mention & relevant intersectional variables?\\
    & Does the document mention & intentionally excluded variables?\\
    & Does the document mention & a rationale for including vulnerable, minority or protected groups?\\
    & Does the document mention & safeguards to protect vulnerable, minority or protected groups?\\ \midrule
    \multirow{7}{*}{Compensation} & Does the document mention & compensation for participation in data work?\\
    & Does the document mention if & compensation is benchmarked?\\
    \textbf{Justice} & Does the document mention if & training time is compensated?\\
    & Does the document mention & minimum withdrawal amounts of compensation?\\
    & Does the document mention & transaction fees to withdraw compensation?\\
    & Does the document mention & the unit for which payment is made?\\
    & Does the document mention & the frequency of payment of data workers?\\ \midrule
    \multirow{5}{*}{Recruitment} & Does the document mention & data subject selection criteria?\\
    \textbf{Justice} & Does the document mention & protocols for verifying vendors?\\
    & Does the document mention & a process for reviewing vendors?\\
    & Does the document mention & a process for onboarding vendors?\\
    & Does the document mention & a process for holding vendors accountable?\\
    & Does the document mention & protocols for verifying subcontractor labour standards?\\ \midrule
    & Does the document mention if & task allocation is transparent?\\
    & Does the document mention & training and guidelines for data workers?\\
    \textbf{Justice} & Does the document mention & the maximum daily working hours for data workers?\\
    Participation & Does the document mention & automated harmful content checks?\\
    Conditions & Does the document mention & occupational health and safety checks?\\
    & Does the document mention & assessments of mental health risks?\\
    & Does the document mention & protections against mental health risks?\\
    & Does the document mention & permission of unionization and collective bargaining for data workers?\\ \midrule 
    \multirow{7}{*}{Fair Treatment} & Does the document mention if & full task details were disclosed to data workers prior to task acceptance?\\
    \textbf{Justice} & Does the document mention if & compensation was disclosed to data workers prior to task acceptance?\\
    & Does the document mention & a baseline estimate of task duration?\\
    & Does the document mention if & task refusal has consequences?\\
    & Does the document mention & compensation for rejected work in absence of quality issues?\\
    & Does the document mention & a mechanism to facilitate complaints of data workers?\\
    & Does the document mention if & data workers can appeal unfair decisions?\\ \midrule
    & Does the document mention & the proportion of harmful content?\\
    & Does the document mention & which harmful content is included/excluded?\\
    \textbf{No-Harm} & Does the document explain why & the inclusion of harmful content is justified?\\
    Harmful & Does the document mention how & harmful content was identified?\\
    Content & Does the document mention how & harmful content was verified?\\
    & Does the document mention & an uncertainty estimate for identified harmful content?\\
    & Does the document mention if & harmful content can be flagged during annotation?\\ \midrule
    \multirow{9}{*}{Privacy} & Does the document mention & the proportion of privacy sensitive content?\\
    & Does the document mention & which privacy sensitive content is included/excluded?\\
    & Does the document explain why & the inclusion of privacy sensitive content is justified?\\
    \textbf{No-Harm} & Does the document mention how & metadata was anonymized or pseudonymized?\\
    & Does the document mention how & privacy sensitive content was identified?\\
    & Does the document mention how & privacy sensitive content was verified?\\
    & Does the document mention & an uncertainty estimate for identified privacy sensitive content?\\
    & Does the document mention & specific data protection regulations that were complied with?\\
    & Does the document mention if & data collection is limited to necessary data?\\ \midrule
    & Does the document mention if & the dataset has a license?\\
    \textbf{No-Harm} & Does the document mention if & the dataset complies with upstream licenses?\\
    Ownership, & Does the document mention & the license of individual data records?\\
    Licensing \& & Does the document mention if & third-party IP is redacted?\\ 
    Attribution & Does the document mention if & copyrighted data instances are excluded? \\
    & Does the document mention if & licenses of data instances permit AI training? \\ \midrule
    & Does the document mention & a specific purpose for data collection? \\
    & Does the document mention & limitations of use for the dataset? \\
    & Does the document mention & a justification for data collection?\\
    & Does the document mention & target variables?\\
    & Does the document mention & the authors?\\
    \textbf{Transparency} & Does the document mention & project sponsors?\\
    Provenance & Does the document mention & the origin / source of data records?\\
    & Does the document mention & data collection methods?\\
    & Does the document mention how & data is collected?\\
    & Does the document mention & involvement of third parties?\\
    & Does the document mention & a digital object identifier for the dataset?\\
    & Does the document mention & documentation? \\ \midrule
    & Does the document mention & variable definitions for metadata?\\
    & Does the document mention & variable data types for metadata?\\
    & Does the document mention & definitions for all variable value options?\\
    & Does the document mention & personal and sensitive variables?\\
    \textbf{Transparency} & Does the document mention & environment variables?\\
    Metadata & Does the document mention & instrument variables?\\
    & Does the document mention & measurable variables?\\
    & Does the document mention & observable variables?\\
    & Does the document mention & subjective variables?\\
    & Does the document mention & inferred variables?\\ \midrule
    & Does the document mention & the annotation method(s)?\\
    & Does the document mention & the annotation protocol / guidelines?\\
    & Does the document mention & the annotation platform(s)?\\
    & Does the document mention & automated tools for annotation?\\
    & Does the document mention & skills criteria for human annotators?\\
    \textbf{Transparency} & Does the document mention & training and guidelines for human annotators?\\
    Annotations & Does the document mention & multiple human annotations for subjective observations?\\
    & Does the document mention & annotator demographics?\\
    & Does the document mention & attention checksfor annotators?\\
    & Does the document mention & human annotator feedback?\\
    & Does the document mention & interannotator agreement?\\
    & Does the document mention & disagreement resolution?\\ \midrule
    & Does the document mention & an evaluation of tasks?\\
    \textbf{Reliability} & Does the document mention & cross-dataset generalization?\\
    Utility & Does the document mention & variance due to performative, non-personal variables?\\
    & Does the document mention & prediction confidence?\\ \midrule
    & Does the document mention & specific variables of interest for diversity?\\
    \textbf{Reliability} & Does the document mention & descriptive statistics?\\
    Dataset & Does the document mention & measurement of dataset diversity?\\
    Measurements & Does the document mention & disaggregated measures across salient groups?\\ \midrule
    & Does the document mention & a quality review protocol?\\
    & Does the document mention & the removal of duplicates?\\
    \textbf{Reliability} & Does the document mention & guidelines for validating annotations?\\
    Quality & Does the document mention if & data records have been manually reviewed?\\
    Control & Does the document mention & a process for issue resolution?\\
    & Does the document mention & annotation consistency checks?\\
    & Does the document mention & annotation reliability checks?\\ \midrule
    & Does the document mention & a process for data record removal? \\
    \textbf{Reliability} & Does the document mention & dataset versioning?\\
    Maintenance & Does the document mention & dataset deprecation?\\ \midrule
    \textbf{Accountability} & Does the document mention & an independent ethics review?\\
    Audit & Does the document mention & a dataset audit?\\ \midrule
    \textbf{Accountability} & Does the document mention & implementation of external standards? \\
    Standards & Does the document mention & certification of compliance with standards?\\ \midrule
\end{longtable}
\end{scriptsize}

\newpage
\subsection{Supporting Material: Multimodal Datasets}
\label{a:multimodal_datasets}

\begin{figure}[hbt]
    \centering
    \includegraphics[width=\linewidth]{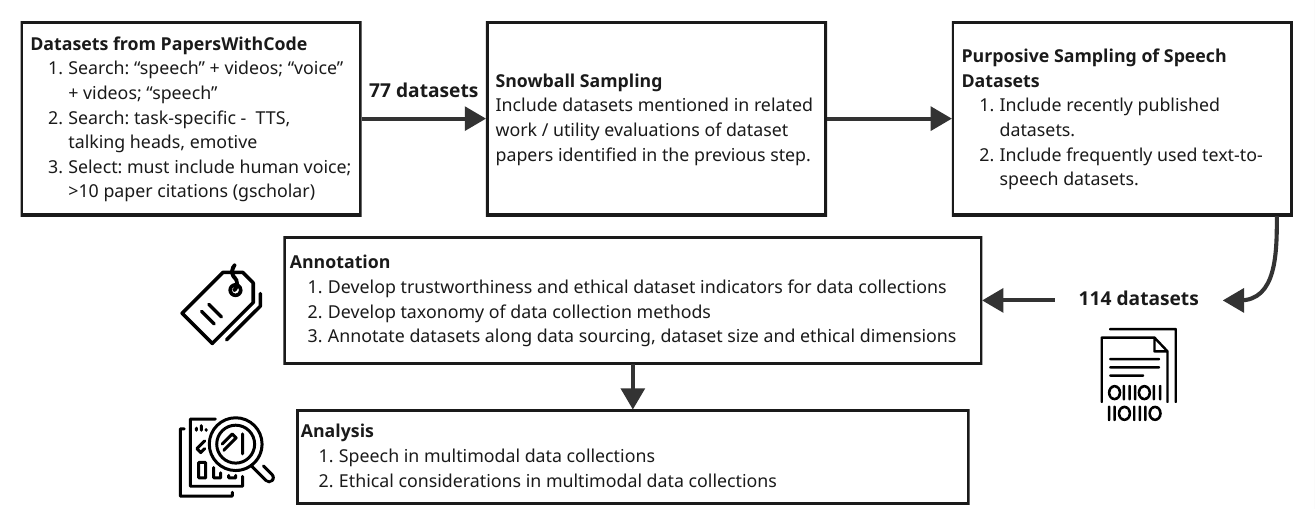}
    \caption{Research approach of this study, including dataset selection, annotation, and analysis.}
    \label{fig:method}
\end{figure}

\begin{figure}[hbt]
    \centering
    \includegraphics[width=0.49\linewidth]{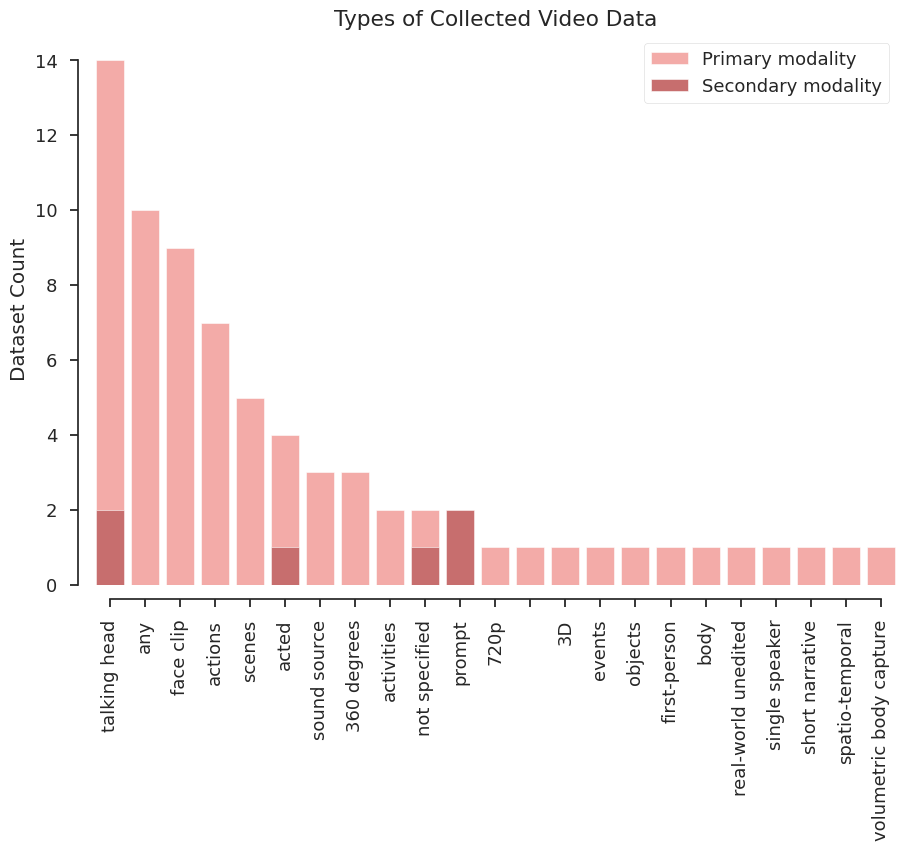}
    \includegraphics[width=0.49\linewidth]{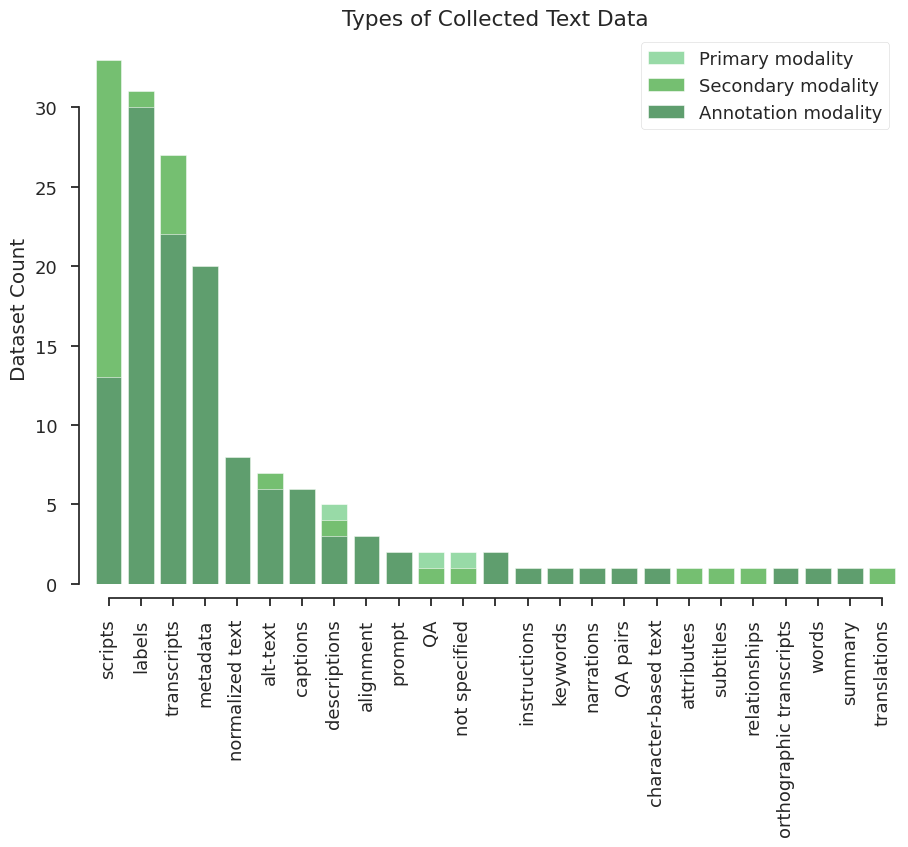}
    \includegraphics[width=0.55\linewidth]{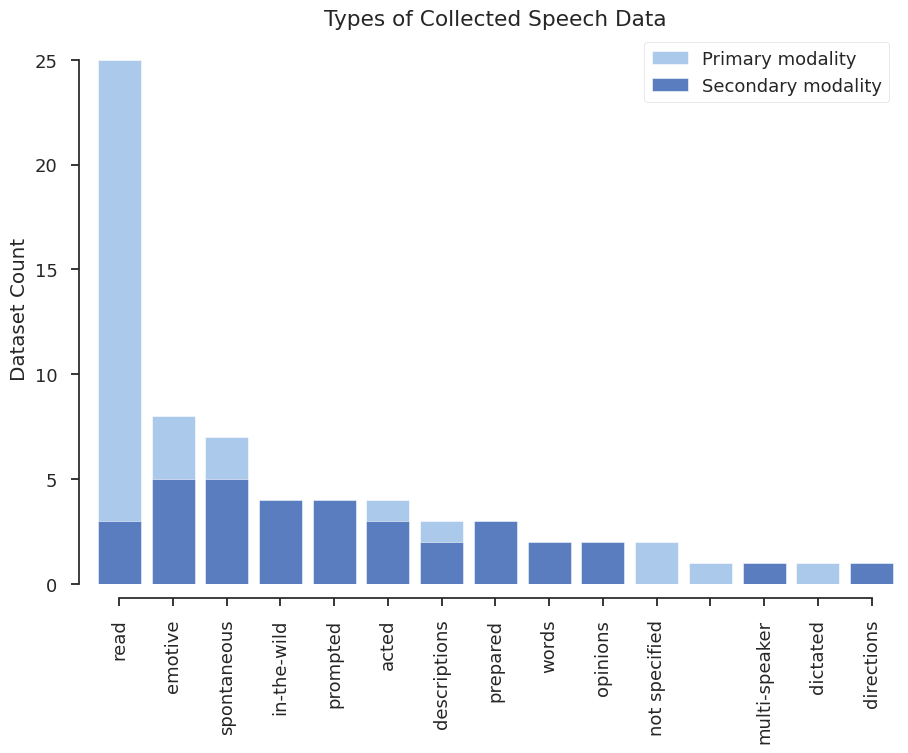}
    \caption{Histograms of detailed data types for video, text and speech for primary, secondary and annotation modalities}
    \label{fig:hist_data_detail}
\end{figure}

\begin{figure}[hbt]
    \centering
    \includegraphics[width=0.9\linewidth]{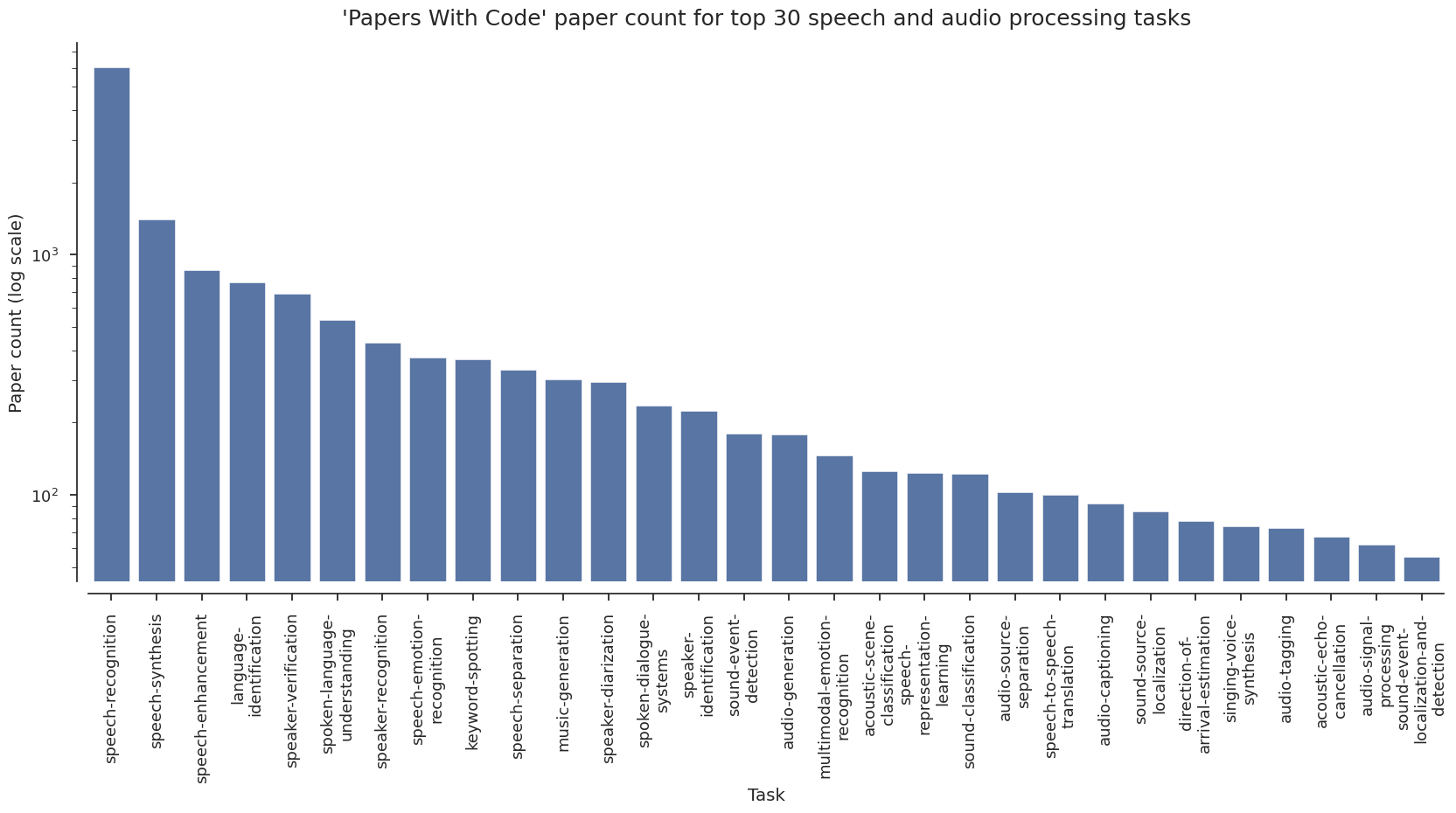}
    \caption{Count of papers per speech processing task on PapersWithCode}
    \label{fig:pwc_paper_count}
\end{figure}

\begin{figure}[hbt]
    \centering
    \includegraphics[width=0.9\linewidth]{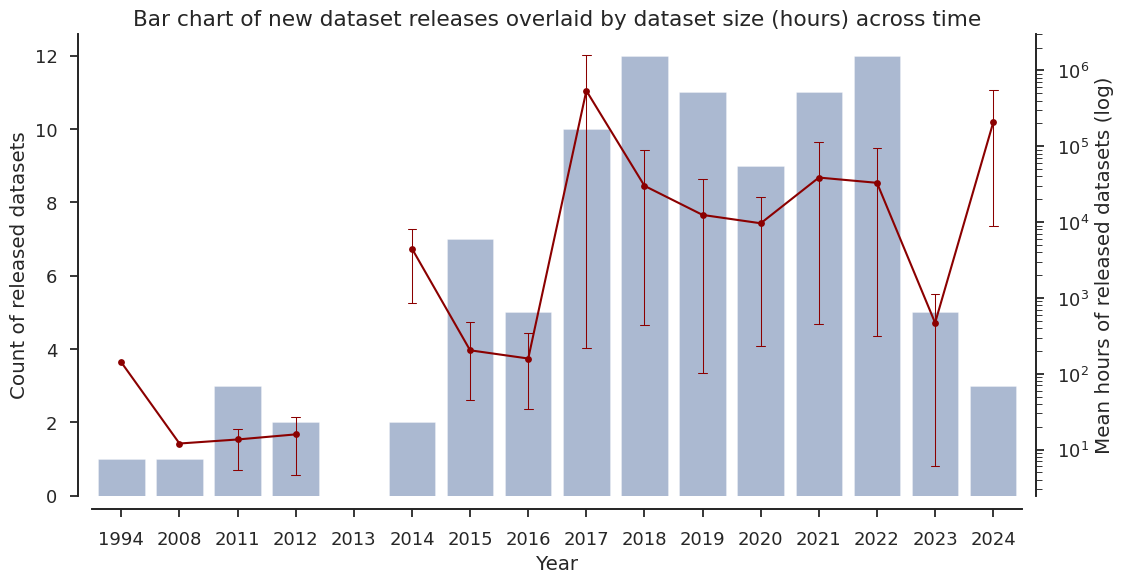}
    \caption{New dataset releases (bar chart) overlaid by mean dataset size (hours) over time, with error bars indicating 95\% confidence intervals. For datasets where the duration in hours was not stated, but where information about the total number of video clips was provided, we assumed a clip duration of 10s to approximate the dataset size. }
    \label{fig:release_count_hours}
\end{figure}

\begin{table}[hbt]
\setlength{\tabcolsep}{4pt}
  \caption{Taxonomy of Data Collection Methods and Example Datasets}
  \label{tab:data_collection_tax_long}
  \centering
  \scriptsize
  \renewcommand{\arraystretch}{1.25}
  \begin{tabular}{L{0.14\textwidth}L{0.15\textwidth}L{0.5\textwidth}L{0.09\textwidth}}
    \toprule
\textbf{Collection Method} & \textbf{Sub-category} & \textbf{Definition} & \textbf{Example} \\ \midrule
\multirow{2}{*}{\textbf{sampling}} & convenience & Data samples selected that are easy to access. & YouCook2 \\
 & purposive & Data samples selected based on implicit or explicit rules, typically based on a researcher's judgement and a purpose they seek to accomplish. & DAVIS \\ \midrule
\multirow{6}{*}{\textbf{direct collection}} & self-reported & Annotations collected via self-assessment by the data subject. & CCV2 \\
 & expert & Annotator has a high degree of expertise and significant training in the domain. & HMDB51 \\
 & amateur & Annotator has limited expertise and training in the domain. & MUStARD \\
 & lab & Data samples are collected in a lab or studio environment controlled by the data collectors. & CREMA-D \\
 & wild & Data samples are collected in an unconstrained, typically public environment. & MISP2021\\
 & constrained & Data samples are collected in a constrained environment controlled by the data subject. & RoomReader \\ \midrule
\multirow{3}{*}{\textbf{proprietary}} & internal & Proprietary data samples from the data collector's or custodian's internal collection (not necessarily public). & 3MASSIV \\
 & agreement & Data samples acquired from a proprietary source through a formal agreement. & LRS2 \\
 & purchased & Data samples purchased from a proprietary source. & MPII-MD\\ \midrule
\multirow{3}{*}{\textbf{crowdsourced}} & volunteer & Participants in the crowdsourced data collection did so on a voluntary basis. & Common Voice \\
 & compensated & Participants in the crowdsourced data collection got incentivised through non-financial compensation for their contribution. & RAVDESS (ann.)\\
 & paid & Participants in the crowdsourced data collection got paid for their contribution. & Charades \\ \midrule
\multirow{3}{*}{\textbf{scraped}} & crawled & Data scraped from the web, indiscriminate of source (i.e. who uploaded) and topic. & Youtube-100M\\
 & sources with curation & Sources with editing, review and publishing processes (e.g. a website like Wikipedia, or a YouTube channel like TED), indiscriminate topic. & VoxPopuli \\
 & constrained & Indiscriminate source, selected topics/people/media types only. & Audioset \\ \midrule
\multirow{6}{*}{\textbf{derived}} & extracted & Extracting secondary or metadata of an existing data collection without modification. & VoxCeleb2 (sec.)\\
 & copy & Selecting an existing data collection in its entirety. & Libryheavy \\
 & subset & Selecting a subset of data samples or annotations from an existing collection, typically based on some filtering criteria. & Librispeech \\
 & expanded & Adding new primary data samples to an existing data collection, or creating a new collection from a seed dataset. & \\
 & extended & Adding new secondary modalities or annotations to an existing or new data collection. & AIShell-1 (sec.)\\
 & composition & Combining data samples in an existing data collection, or from different data collections. & Visual-Genome \\ \midrule
\multirow{2}{*}{\textbf{synthetic}} & expanded & Adding new primary data samples, or creating a new collection from a seed dataset using synthetic or automatic means. &  AV-DeepFake1M\\
 & extended & Adding new secondary modalities or annotations to an existing or new collection using synthetic or automatic means. & LAION-400M (sec.) \\
    \bottomrule
  \end{tabular}
\end{table}

\newpage


\begin{figure}[hbt]
    \centering
    \includegraphics[angle=90, width=0.9\linewidth]{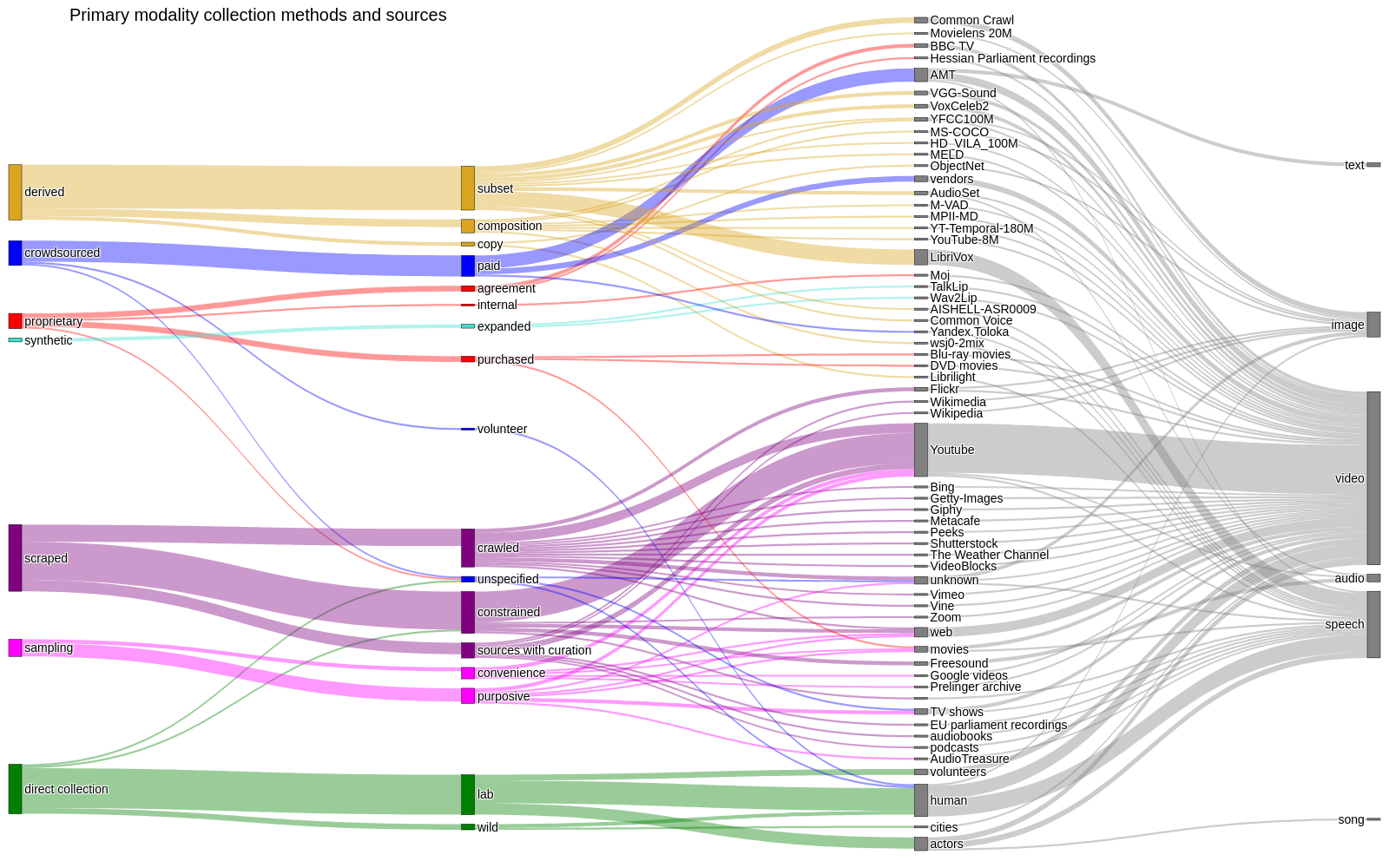}
    \caption{Primary data collection methods (highlighted in colour) and sources}
    \label{fig:sankey_primary}
\end{figure}

\begin{figure}[hbt]
    \centering
    \includegraphics[angle=90, width=0.9\linewidth]{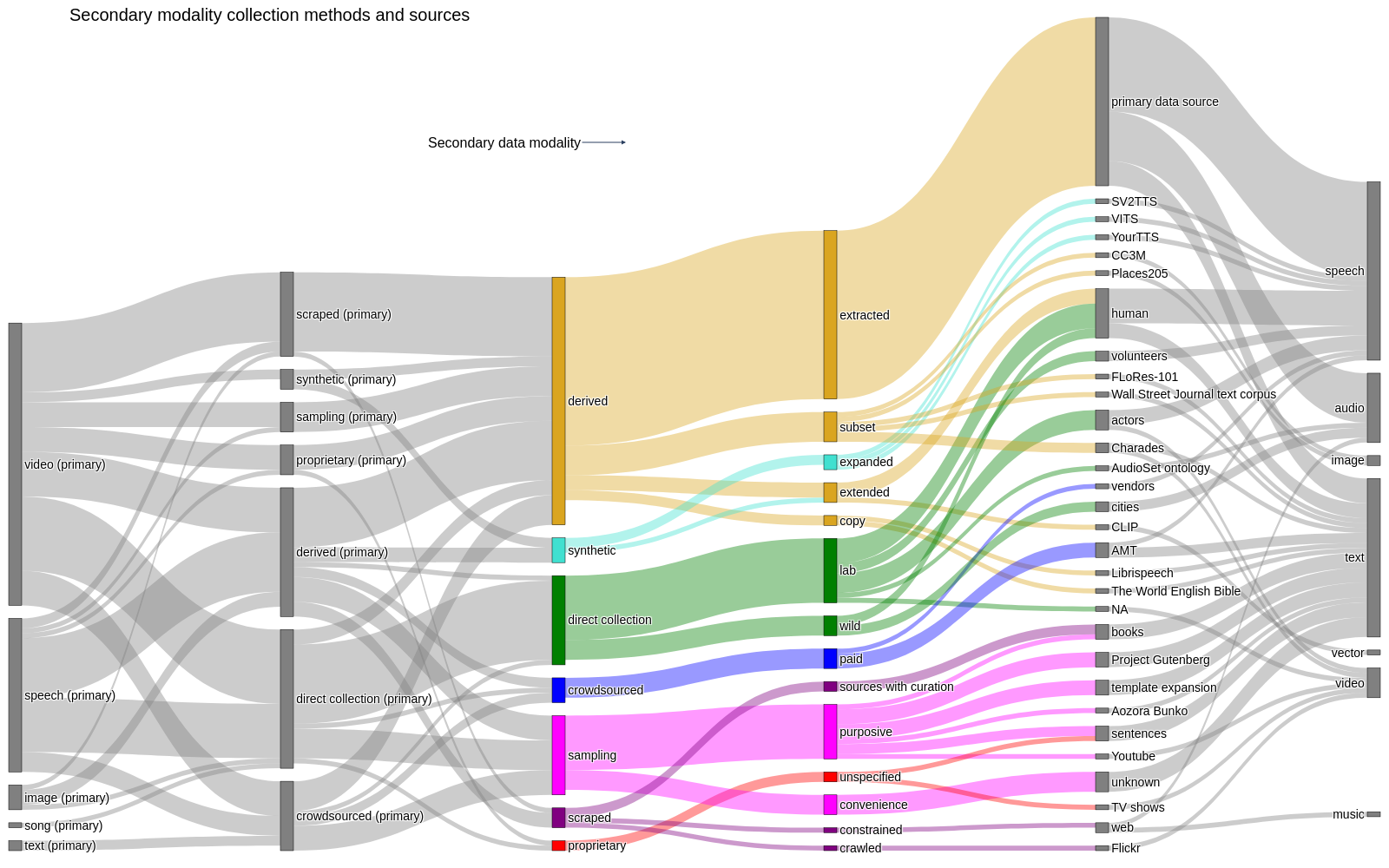}
    \caption{Secondary data collection methods (highlighted in colour) and sources}
    \label{fig:sankey_secondary}
\end{figure}

\begin{figure}[hbt]
    \centering
    \includegraphics[angle=90, width=0.9\linewidth]{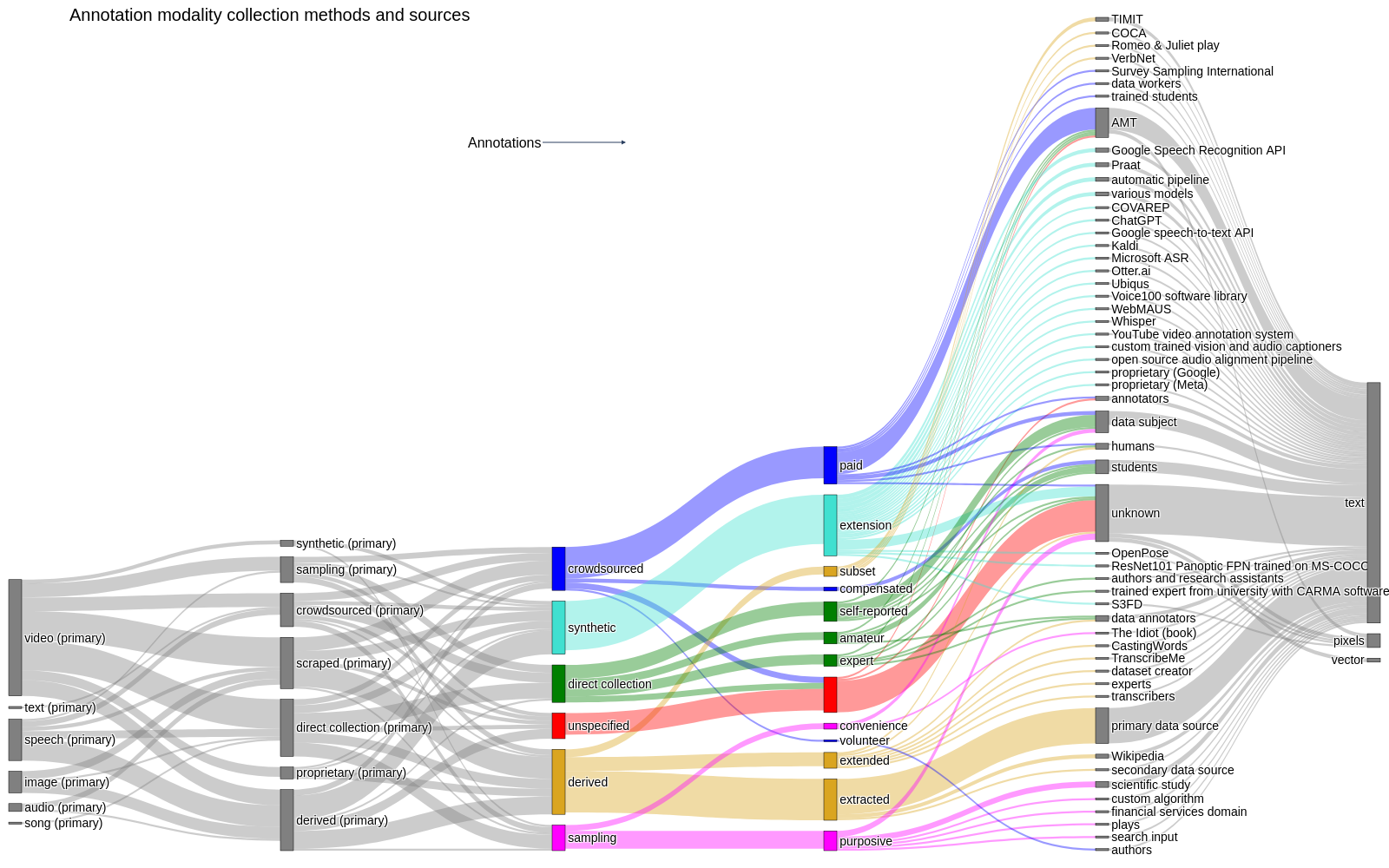}
    \caption{Annotation data collection methods (highlighted in colour) and sources}
    \label{fig:sankey_annotation}
\end{figure}

\newpage




\end{document}